\documentstyle[12pt]{article}
\baselineskip=\normalbaselineskip
\ifx\TwoupWrites\UnDeFiNeD\else\target{\magstepminus1}{11.3in}{8.27in}
	\source{\magstep0}{7.5in}{11.69in}\fi
\newfont{\fourteencp}{cmcsc10 scaled\magstep2}
\newfont{\titlefont}{cmbx10 scaled\magstep2}
\newfont{\authorfont}{cmcsc10 scaled\magstep1}
\newfont{\fourteenmib}{cmmib10 scaled\magstep2}
	\skewchar\fourteenmib='177
\newfont{\elevenmib}{cmmib10 scaled\magstephalf}
	\skewchar\elevenmib='177
\makeatletter
\newcommand\nonsequentialeqnum{
	\@addtoreset{equation}{section}
	\def\theequation{\arabic{section}.\arabic{equation}}}
\newif\ifp@bblock  \p@bblocktrue
\newcommand\nopubblock{\p@bblockfalse}
\newcommand\topspace{\hrule height 0pt depth 0pt \vskip}
\newcommand\p@bblock{\begingroup \tabskip=\hsize minus \hsize
	\baselineskip=1.5\ht\strutbox \topspace-2\baselineskip
	\halign to\hsize{\strut ##\hfil\tabskip=0pt\crcr
	\the\Pubnum\crcr\the\date\crcr}\endgroup}
\newcommand\YITPmark{\hbox{\fourteenmib YITP\hskip0.2cm
        \elevenmib Uji\hskip0.15cm Research\hskip0.15cm Center\hfill}}
\renewcommand\titlepage{\ifx\TwoupWrites\UnDeFiNeD\null\vspace{-1.7cm}\fi
	\YITPmark\vskip0.6cm
	\ifp@bblock\p@bblock \else\hrule height 0pt \relax \fi}
\makeatother
\newtoks\date
\newtoks\Pubnum
\newtoks\pubnum
\Pubnum={YITP/U-\the\pubnum}
\date={\today}
\newcommand{\frontpageskip}{\vspace{12pt plus .5fil minus 2pt}}
\renewcommand{\title}[1]{\frontpageskip
	\begin{center}{\titlefont #1}\end{center}\par}
\renewcommand{\author}[1]{\frontpageskip\par\begin{center}
	{\authorfont #1}\end{center}
	\nobreak
	}

\newcommand{\address}[1]{\par\begin{center}{\sl #1}\end{center}\par}

\renewcommand{\thanks}[1]{\footnote{#1}}
\renewcommand{\abstract}{\par\frontpageskip\centerline{\fourteencp Abstract}
	\vspace{8pt plus 3pt minus 3pt}}
\newcommand\YITP{\address{Uji Research Center \\
	       Yukawa Institute for Theoretical Physics\\
               Kyoto University,~Uji 611,~Japan\\}}
\begin{document}
\thispagestyle{empty}
\newcommand{\bb}{\begin{equation}}
\newcommand{\ee}{\end{equation}}
\renewcommand{\a}{\alpha}
\renewcommand{\pd}{\partial}
%

%\nopubblock        %% uncomment in making submit-version
\nonsequentialeqnum %% uncomment in (Section.Number) equation number style.
\pubnum{93-28\cr hep-th/9309094}
\date{September 1993}
\titlepage

\title{Physical States, Factorization and Nonlinear Structures
in Two Dimensional Quantum Gravity}

\author{Ken-ji HAMADA\footnote{E-mail
address: hamada@yisun1.yukawa.kyoto-u.ac.jp}}

\YITP

\abstract{
The nonlinear structures in 2D quantum gravity coupled to the
$(q+1,q)$ minimal model are studied in the Liouville theory to clarify
the factorization and the physical states. It is confirmed that
the dressed primary states outside the minimal table are identified
with the gravitational descendants. Using the discrete states of
ghost number zero and one we construct the currents and investigate
the Ward identities which are identified with the W and the Virasoro
constraints. As nontrivial examples we derive the $L_0$, $L_1$ and
$W_{-1}^{(3)}$ equations exactly. $L_n$ and $W^{(k)}_n$ equations are
also discussed. We then explicitly show the decoupling of the edge
states $O_j ~(j=0~ {\rm mod}~ q) $. We consider the interaction theory
perturbed by the cosmological constant $O_1$ and the screening charge
$S^+ =O_{2q+1}$. The formalism can be easily generalized to potential
models other than the screening charge.
}

\newpage

\section{Introduction}
\indent

 Since the discovery of the double scaling limit in the matrix
models~\cite{gm,ds,bk},
many efforts have been made to understand 2D quantum gravity. In the
matrix model approach it was shown that the amplitudes obey the
nonlinear equations called the W and the Virasoro
constraints~\cite{fkn,dvv}.
We have tried to investigate the nonlinear structures
in the Liouville theory approach [6--15] in order to clarify the
nature of physical states and factorization properties of amplitudes.
There are a few works~\cite{pb,ha}, but clear arguments have not been
found yet.

  In the Liouville theory  it has been found that there are an
infinite number of the BRST invariant states. Although the
discrete states are  discussed in detail~\cite{lz,bmpa},  the
physical role of them are not sufficiently understood. Here we discuss
the role of the discrete states with ghost number zero and one
according to the classification by Bouwknegt, McCarthy and Pilch (BMP).
We also consider the dressed primary states both inside and outside
the minimal table. As shown by Kitazawa~\cite{kit} the dressed
primary states
outside the minimal table no longer decouple in the combined
matter-Liouville theory. Furthermore he claimed these fields to be the
gravitational descendants. In this paper,  using the discrete states of
the ghost number zero and one, we construct symmetry
currents.\footnote{The importance of the BRST invariant
discrete states of
ghost number zero is emphasized in the two dimensional string
theory~\cite{w,kp,kle}. Also in
ref.~\cite{pb,ha} they play an important role.}
%%%%%%%%%%%%
Substituting the currents into the correlation functions of the
dressed primary fields we set up the Ward identities and identify
them with the W and the Virasoro constraints.

  The paper is organized as follows. In Sect.2 we summarize the
various BRST invariant states of the $(q+1,q)$ minimal theory coupled
to gravity and define the symmetry currents.
The correlation functions are defined along the argument of Goulian
and Li~\cite{gl}. We consider the interaction theory perturbed by the
cosmological constant $O_1$ and the screening charge $S^+ =O_{2q+1}$,
where another screening charge $S^-$ is not used. In Sect.3 we discuss
the factorization properties of amplitudes. The factorization in the
Liouville theory is rather similar with that of the string
theory~\cite{s,pa}.
However there is a crucial difference in the pole structures of
the propagators, which leads to the nonlinear structures in the
Liouville theory approach. In Sect.4 we discuss the Ward
identities corresponding to the Virasoro costraints. As nontrivial
examples we derive $L_0 $ and $L_1$ equations exactly. We also discuss
the $L_n $ equations. The Ward
identities corresponding to the W constraints are discussed in Sect.5,
where $W^{(3)}_{-1}$ equation is derived exactly. $W^{(k)}_n$
equations are briefly discussed. Then we explicitly
show that the edge states $O_j ~ (j=0~ {\rm mod}~ q)$ decouple from the
expressions. Sect.6 is devoted to discussion. Here we argue the
generalization of the formalism.
If we consider the interaction theory perturbed by the scaling
operator $O_{p+q} $ instead of the screening charge $O_{2q+1}$, we
will get the gravity theory coupled to $(p,q)$ conformal matter.
We also  comment on the  unitarity problem in the
two dimensional quantum gravity.

\section{BRST invariant states and correlation functions}
\indent

   In this section we summarize the Liouville theory approach to 2D
quantum gravity coupled to the minimal conformal matter with central
charge $c_M =1-\frac{6}{q(q+1)}$ and establish our notations and
conventions.  The action for the Liouville-matter part is
\begin{eqnarray}
  &&  S_0 =\frac{1}{8\pi}\int \sqrt{\hat g}({\hat g}^{\alpha\beta}
            \partial_{\alpha}\phi \partial_{\beta}\phi
                 +2i Q_L {\hat R}\phi )
              \nonumber \\
  && \qquad  +\frac{1}{8\pi}\int \sqrt{\hat g}({\hat g}^{\alpha\beta}
            \partial_{\alpha}\varphi \partial_{\beta}\varphi
                +2i Q_M {\hat R}\varphi ) ~,
\end{eqnarray}
where the scalar fields $\phi$ and $\varphi$ are the Liouville and the
matter fields respectively. The background charges are
\bb
     iQ_L = \frac{2q+1}{\sqrt{2q(q+1)}} ~ , \qquad
     Q_M = \frac{1}{\sqrt{2q(q+1)}} ~,
\ee
where we use the convention that $Q_L$ is purely imaginary. In terms
of modes
\bb
    i\partial \phi (z)=\sum_{n\in {\bf Z}} \alpha^L_n z^{-n-1} \qquad
    i\partial \varphi (z)=\sum_{n\in {\bf Z}} \alpha^M_n z^{-n-1} ~,
\ee
the Virasoro algebras are given by
\bb
   L^{L,M}_n =\frac{1}{2}\sum_{m \in {\bf Z}}
                 :\alpha^{L,M}_m \alpha^{L,M}_{n-m}:
                   -(n+1)Q_{L,M}\alpha^{L,M}_n ~,
\ee
where $[\alpha^L_n , \alpha^L_m ] =[\alpha^M_n , \alpha^M_m ]=n
\delta_{n+m,0}$.

  The physical states are identified with nontrivial cohomology
classes of the BRST operator
\bb
      Q_{BRST} = \oint \frac{dz}{2\pi i}c(z)\Bigl( T^L (z) +T^M (z)
                      +\frac{1}{2} T^G (z) \Bigr)~.
\ee
In terms of modes
\begin{eqnarray}
  &&   Q_{BRST} =  c_0 L_0 -b_0 M +{\hat d} ~, \\
  &&   L_0 =  L^L_0 +L^M_0 +L^G_0 ~, \qquad
                 M=\sum_{n \neq 0} n :c_{-n}c_n : ~, \\
  &&   {\hat d} = \sum_{n\neq 0}c_{-n} (L^L_n +L^M_n )
                   -\frac{1}{2}\sum_{n,m\neq 0 \atop n+m \neq 0}
                     (m-n):c_{-m} c_{-n} b_{n+m}: ~.
\end{eqnarray}
By introducing the lightcone-like combinations of the modes
$\alpha^{\pm}_n =\frac{1}{\sqrt 2}(\alpha^M_n \pm i\alpha^L_n )$, $n
\neq 0$ and the generalized momentum variables
\bb
     P^{\pm}(n)=\frac{1}{\sqrt 2} \Big[ (\alpha^M_0 -(n+1)Q_M )
                  \pm i(\alpha^L_0 -(n+1)Q_L ) \Big]
\ee
the operator ${\hat d}$ is rewritten as
\begin{eqnarray}
   &&  {\hat d}=\sum_{n\neq 0}c_{-n}(\alpha^-_n P^+ (n)
                          +\alpha^+_n P^-(n))
               \nonumber \\
   && \qquad\qquad  +\sum_{n,m \neq 0 \atop n+m\neq 0}:c_{-n}
                 \Bigl(\alpha^+_{-m}\alpha^-_{n+m}
                     +\frac{1}{2}(m-n)c_{-m}b_{n+m}\Bigr): ~.
\end{eqnarray}

   BMP~\cite{bmpa} classify the nontrivial cohomology of the BRST
charge. If $P^+(r)
\neq 0$ or $P^-(s)\neq 0$ for all $r,s \in {\bf Z},~ r,s \neq 0$ and
 $P^+(0)P^-(0)=0$, then there exist the dressed primary states with
momentums $i\alpha^L_0 =\alpha_j $ and $\alpha^M_0 =\beta_j $
parametrized by
\bb
     \alpha_j =(2q+1-j)Q_M ~, \qquad \beta_j =(1+j)Q_M ~.
\ee
The corresponding fields are given by
\bb
      O_j = \int d^2 z V_j (z,{\bar z})
          =\int d^2 z {\rm e}^{\a_j \phi(z,{\bar z})}
                      {\rm e}^{i\beta_j \varphi(z,{\bar z})} ~.
\ee
Here we use the unusual convention for $\beta_j $. The reason why we
adopt it will be mentioned after  correlation functions are defined.
We identify these fields with the gravitationl primaries and their
descendants as
\bb
        O_{nq+k} =\sigma_n (O_k ) ~,\qquad ( k=1, \cdots ,q-1;
                    \quad n\in {\bf Z}_{\geq 0})~,
\ee
where $n=0$ states are gravitational primaries. We now exclude the edge
states of $k=0 $. This identification was first proposed by
Kitazawa~\cite{kit}, who
showed that the dressed primary states outside the minimal table fail
to decouple in the combined matter-Liouville theory. In the following
section, by deriving the nonlinear structures directly in the
Liouville theory approach, we will confirm this identification and
show that the edge states indeed decouple.

   Besides the dressed primary states there exist the nontrivial BRST
invariant states called the discrete states at the
momentums~\footnote{Note that $\beta_{r,s}=\beta_{(r-s)q+r}$, where
$\beta_{r,s}=\frac{1}{\sqrt{2}}(1+r)\beta_+
+\frac{1}{\sqrt{2}}(1+s)\beta_-$, $\beta_+ =\sqrt{\frac{q+1}{q}}$ and
$\beta_- =-\sqrt{\frac{q}{q+1}}$.}
%%%%%%%%%%%%%%
\bb
     i\a^L_0 =\a_{-(r+s)q-r} ~, \qquad \a^M_0 =\beta_{(r-s)q+r} ~.
\ee
In this paper we only cosider the discrete states of $r,s<0 $. Then
there exist the states of ghost number zero, $B_{r,s}(z)$. For examples
we get
\begin{eqnarray}
 &&  B_{-1,-1}(z) =  1   ~,        \\
 &&  B_{-1,-2}(z) =  \biggl[ bc-\sqrt{\frac{q+1}{2q}}
                    (\partial \phi +i\partial\varphi) \biggr]
                     {\rm e}^{\a_{3q+1}\phi(z)}
                       {\rm e}^{i\beta_{q-1}\varphi(z)}   ~,        \\
 &&  B_{-2,-1}(z) =  \biggl[ bc-\sqrt{\frac{q}{2(q+1)}}
                    (\partial \phi -i\partial\varphi) \biggr]
                     {\rm e}^{\a_{3q+2}\phi(z)}
                       {\rm e}^{i\beta_{-q-2}\varphi(z)}   ~.
\end{eqnarray}
There also exist the primary states $R_{r,s}$ with the momentums
(2.14). For instance
\begin{eqnarray}
 && R_{-1,-1}(z)= \pd \phi(z) +  (2q+1) i \pd \varphi(z) ~,     \\
 && R_{-1,-2}(z)= \frac{1}{q}\biggl[
                -\frac{q-3}{4\sqrt{2}}  \pd^2 \phi
                +\frac{7q+3}{4\sqrt{2}} i\pd^2 \varphi
                +\frac{q^2 -2q-1}{8\sqrt{q(q+1)}}
                      \Bigl\{ (\pd \phi)^2 -(\pd \varphi)^2 \Bigr\}
                       \biggr. \nonumber  \\
 &&               \qquad\qquad\qquad\qquad\qquad \biggl.
                    -\frac{3q^2+2q+1}{4\sqrt{q(q+1)}} \pd\phi i\pd\varphi
                 \biggr]  {\rm e}^{\a_{3q+1}\phi(z)}
                       {\rm e}^{i\beta_{q-1}\varphi(z)}   ~,            \\
 && R_{-2,-1}(z)= \frac{1}{q+1}\biggl[
                \frac{q+4}{4\sqrt{2}}  \pd^2 \phi
                +\frac{7q+4}{4\sqrt{2}} i\pd^2 \varphi
                -\frac{q^2 +4q+2}{8\sqrt{q(q+1)}}
                      \Bigl\{ (\pd \phi)^2 -(\pd \varphi)^2 \Bigr\}
                           \biggr. \nonumber   \\
 &&               \qquad\qquad\qquad\qquad\qquad\qquad \biggl.
                    -\frac{3q^2+4q+2}{4\sqrt{q(q+1)}} \pd\phi i\pd\varphi
                 \biggr]  {\rm e}^{\a_{3q+2}\phi(z)}
                       {\rm e}^{i\beta_{-q-2}\varphi(z)}   ~,
\end{eqnarray}
where we remove $c$ ghost in the definition of $R_{r,s}$.

   Combining $R_{r,s}$ and ${\bar B}_{r,s}$ we can construct the symmetry
currents
\bb
    W_{r,s}(z,{\bar z})=R_{r,s}(z){\bar B}_{r,s}({\bar z}) ~,
                \qquad r,s \in {\bf Z}_- ~,
\ee
which satisfy
\bb
     \pd_{{\bar z}} W_{r,s}(z,{\bar z})
            =\{ {\bar Q}_{BRST} ,[{\bar b}_{-1},
                              W_{r,s}(z,{\bar z})] \} ~.
\ee
Main assertion of this paper is that the Ward identities of the currents
\bb
       \int d^2 z \partial_{\bar z}
         \ll W_{-k,-n-k}(z,{\bar z})\prod_{j\in S} O_j \gg_g =0 ~,
           \quad (k=1, \cdots q-1;~ n \in {\bf Z}_{\geq 2})
\ee
are just the $W^{(k+1)}_{n}$ constraints. The equations for $k=1$ is
the Virasoro constraints and others are the $W$ constraints.

  Let us define the correlation functions of the Liouville theory. We
consider the interaction theory,
\bb
       S=S_0 +\mu O_1 -t O_{2q+1} ~,
\ee
where $O_1 $ is the cosmological constant and $O_{2q+1}$ is nothing
but the screening charge $S^+ ={\rm e}^{i\sqrt{2}\beta_+ \varphi}$,
$\beta_+ =\sqrt{(q+1)/q}$. Note that we do not use another
screening charge  $S^- ={\rm e}^{i\sqrt{2}\beta_- \varphi}$,
$\beta_- =-\sqrt{q/(q+1)}$ because it is not included in the
definition of the scaling operators (2.13).

  The matter sector has the symmetry under the constant shift
$\varphi \rightarrow \varphi + 2\pi /Q_M $ because then the
action only shifts by $2\pi i\chi $, where $\chi $ is the Euler number,
such that ${\rm e}^{-S} $ is invariant. Therefore we restrict the range of
the zero mode integral of $\varphi $ within $0 \leq \varphi_0 \leq
2\pi /Q_M $. On the other hand, since $Q_L $ is purely imaginary,
there is no such a symmetry for the Liouville sector so that we do not
restrict the range of zero mode of $\phi $. After integrating over the
zero modes of the Liouville and the matter fields the correlation
functions are expressed as
\bb
   \ll \prod_{j \in S}O_j \gg_g
      =\kappa^{-\chi} \mu^s \frac{\Gamma(-s)}{\a_1}
                \frac{2\pi}{Q_M} \frac{t^n}{n!}
          < \prod_{j \in S}O_j ~ (O_1)^s (O_{2q+1})^n >_g ~,
\ee
where $g $ is genus, $\chi =2-2g$ and
\begin{eqnarray}
    & & s = \frac{1}{\a_1}[iQ_L \chi -\sum_{j \in S}\a_j ]
           = \frac{1}{2q}[(2q+1)\chi -\sum_{j \in S}(2q+1-j) ] ~, \\
    & & n =  \frac{1}{\beta_{2q+1}}[Q_M \chi -\sum_{j \in S}\beta_j
                                       -\beta_1 s]
           = \frac{1}{2q}[-\chi +\sum_{j \in S}(1-j) ] ~.
\end{eqnarray}
We introduce the string coupling constant $\kappa $. The
$\Gamma$-function comes from the zero mode integral of $\phi$. The
$\varphi_0 $ integral gives the Kronecker delta multiplied by
$2\pi /Q_M $ which guarantees the momentum neutrality of matter sector.
The expression connects between the correlators in the interaction
picture $\ll \cdots \gg_g $
and ones in the free picture $< \cdots >_g $. If $s $ and $n$ are
integers, the correlation functions can be
calculated. However $s$ and $n$ are not integers in general. According
to the argument of Goulian and Li~\cite{gl} we define the correlators by
analytic continuations in $s $ and $n$. Then $n!$ is defined by
$\Gamma (n+1)$.

  In the free picture correlators the Liouville field satisfies the
equation of motion ${\bar \pd}\pd \phi =0$. Then we get the following
Ward identity for the Liouville sector,
\begin{eqnarray}
   && {\bar \pd}  \ll  \pd \phi(z) \prod_{j \in S}
                 V_j (z_j,{\bar z}_j) \gg^{(L)}_g   \nonumber    \\
   &&= -\pi \a_1 s \mu^s \frac{\Gamma(-s)}{\a_1}
                 < V_1 (z,{\bar z}) \prod_{j \in S}
             V_j (z_j,{\bar z}_j) (O_1)^{s-1} >^{(L)}_g  \nonumber \\
   &&\quad   -\pi \sum_{k\in S}\a_k \delta^2(z-z_k)
                  \mu^s \frac{\Gamma(-s)}{\a_1}
                      <  \prod_{j \in S}
                   V_j (z_j,{\bar z}_j) (O_1)^s >^{(L)}_g  \\
   && = \pi \a_1  \mu
                \ll V_1 (z,{\bar z}) \prod_{j \in S}
                 V_j (z_j,{\bar z}_j) \gg^{(L)}_g  \nonumber \\
   &&\quad    -\pi \sum_{k\in S}\a_k \delta^2(z-z_k)
                  \ll  \prod_{j \in S}
                  V_j (z_j,{\bar z}_j) \gg^{(L)}_g  ~, \nonumber
\end{eqnarray}
where we use the operator product
\bb
        \pd \phi (z){\rm e}^{\a\phi(w,{\bar w})}
               =-\frac{\a}{z-w}{\rm e}^{\a\phi(w,{\bar w})}
\ee
in the free picture and ${\bar \pd}\frac{1}{z-w}=\pi \delta^2 (z-w)$.
In the second equality the relation $-s\Gamma(-s)=\Gamma(1-s)$ is
used. This expression indicates that in the correlator of the
interaction picture the Liouville field satisfies the equation of
motion
\bb
     {\bar \pd}\pd \phi (z,{\bar z}) =\pi \a_1 \mu V_1 (z,{\bar z}) ~.
\ee
For the matter sector also the same argument succeeds.

  Finally we comment on the convention of the matter momentum
$\beta_j $. If we adopt the convention $\beta_j =(1-j)Q_M $ and the
theory is perturbed by $S^-$, we meet the divergences when the correlation
 functions include the fields of $j=n(q+1),~n\in {\bf Z}_+ $. So we
need the regularization. On the other hand, in our convention,
such a divergence do not appear and then we can explicitly show  the
decoupling of the edge states $O_j ~(j=0~{\rm mod}~q)$.

\section{Factorization properties of amplitudes}
\indent

   The structures of factorization in 2D quantum gravity are rather
similar to the string theory~\cite{s,pa}. However, in a foundamental
point, they are different. The difference leads to the nonlinear
structures of 2D quantum gravity. In the following we develop the argument
comparing the amplitudes of the string theory and 2D quantum gravity.

   Amplitudes in the string theory can be decomposed into vertex
operators and propagators
\bb
      D=\int_{|z|\leq 1}\frac{d^2 z}{|z|^2}
                z^{L_0}{\bar z}^{{\bar L}_0}
       =2\pi \biggl( \frac{1}{H}-\lim_{\tau \rightarrow \infty}
          \frac{1}{H}{\rm e}^{-\tau H} \biggr) ~,
\ee
where $H=L_0 +{\bar L}_0 $. The last term stands for the boundary of
moduli space pinching 2D surface. The intermediate states are
expanded by the normalizable
off-shell eigenstates of the Hamiltonian $H $. The propagator $1/H $
describes the propagation of the off-shell modes.

   In the Liouville theory the intermediate states are also expanded
by the normalizable eigenstates of $H$,
\bb
     H |p,\beta_k ,N > = \Bigl( p^2 -Q^2_L +2(\Delta_k +N-1) \Bigr)
                            |p, \beta_k ,N >    ~,
\ee
where $p $ is real. $\Delta_k $ is the conformal dimension of matter
sector,
\bb
      \Delta_k = \frac{k^2 -1}{4q(q+1)} ~.
\ee
The integer $N$ stands for the oscillation level of the states. The zero
level states are defined by
\bb
     |p, \beta_k > = {\rm e}^{i(p+Q_L)\phi(0)}
                     {\rm e}^{i\beta_k \varphi(0)}
                      |0>_{L,M} \otimes {\bar c}_1 c_1 |0>_G  ~.
\ee
We take the following normalization,
\bb
    \ll p^{\prime},\beta_{k^{\prime}},N^{\prime}
       | p,\beta_k ,N \gg_{g=0}
          = \kappa^{-2} 2\pi\delta(p+p^{\prime})
             \frac{2\pi}{Q_M}\delta_{k+k^{\prime},0}
             \delta_{N,N^{\prime}} ~.
\ee
The zero mode integral of the Liouville field now produces the
$\delta$-function.

  Let us consider a channel of factorization of correlators (2.25)
devided into  sets $F_1$ and $F_2 $ composed of the operators in $S$.
By inserting the complete set we get the expression
\begin{eqnarray}
  & & \kappa^2 \frac{Q_M}{2\pi} \sum^{\infty}_{N=0} \sum^{\infty}_{k=1}
      \int^{\infty}_{-\infty} \frac{dp}{2\pi}
       \ll F_1 |-p, \beta_{-k}, N \gg      \nonumber \\
  & &\qquad \times \frac{2\pi}{p^2 + E_{k,N}} \Bigl(
           1-\lim_{\tau \rightarrow \infty}
                   {\rm e}^{-\tau(p^2 +E_{k.N})} \Bigr)
       \ll p,\beta_k , N | F_2 \gg  ~,
\end{eqnarray}
where $E_{k,N}=-Q^2_L +2(\Delta_k +N-1) $. The cosmological constants
and the screening charges are properly factorized. Note that, since
$E_{k,N}$ is always positive, the pole of the Liouville momentum $p$
lays on the imaginary axis. Therefore the $p$ integral for $1/H$ part
can be deformed to the complex plane and picks up only the
on-shell $(H=0)$ poles on the imaginary axis. This indicates that in
the string theory the intermediate states are off-shell so that the
boundary term vanishes in the limit $\tau \rightarrow \infty $, while
in the 2D quantum gravity the intermediate states becomes on-shell and
so we can not always ignore the boundary term. In fact it plays an important
role when the correlators include the currents $W_{r,s}$.

\section{Virasoro equations}
\indent

  The Virasoro constraints derived from the matrix model are described
symbolically as
\bb
    L_n =\frac{1}{2\lambda} \sum_{-k-l=nq} k l x_k x_l
              +\sum_{-k+m=nq} k x_k \pd_m
           +\frac{\lambda}{2} \sum_{k+l=nq} \pd_k \pd_l   ~,
\ee
where $x_{mq}$ and $\pd_{mq},~ m\in {\bf Z}_{\geq 0}$ are discarded.
The aim of this section is to derive the Virasoro constraints as the
Ward identities of the currents $W_{-1,-n-1}$.

\subsection{$L_0$ equation}
\indent

   Let us discuss the Ward identity for the current $W_{-1,-1}$. This
is rather trivial. As is easily understood from the equation (2.28) it
is equivalent to the momentum neutrality conditions which are,
by definition (2.25), expressed
as~\footnote{From eq.(4.2) we can derive the normalization
independent ratio
$$
    \frac{\ll O_{2q+1} O_k O_k \gg^2_0 \ll 1 \gg_0}{\ll O_{2q+1}
           O_{2q+1}\gg_0 \ll O_k O_k \gg^2_0 }
      =\frac{k^2}{q+1} ~,
$$
which agree with the result derived in ref.~\cite{kit}.}
%%%%%%%%%%%
\begin{eqnarray}
    && t \ll O_{2q+1} \prod_{j\in S} O_j \gg_g
             = n \ll \prod_{j\in S} O_j \gg_g    ~,     \\
    && -\mu \ll O_1 \prod_{j\in S} O_j \gg_g
             = s \ll \prod_{j\in S} O_j \gg_g  ~,
\end{eqnarray}
where $s,n$ are given by (2.26) and (2.27). Combining these equations we get
\bb
    (2q+1)t \ll O_{2q+1}\prod_{j\in S} O_j \gg_g
             + x \ll O_1 \prod_{j\in S} O_j \gg_g
             + \Bigl( \sum_{j\in S} j \Bigr)
                 \ll  \prod_{j\in S} O_j \gg_g =0  ~,
\ee
where $x=-\mu $. This is nothing but the $L_0$ equation
\bb
      L_0 = \sum_k k x_k \pd_k
\ee
with $x_1 =x $, $x_{2q+1}=t$ and other $x_j ~'{\rm s}=0$.

\subsection{$L_1$ equation}
\indent

   The first nontrivial example is the Ward identity for the current
$W_{-1,-2}$. The operator product expansion (OPE) between the current
and the scaling operator is given by
\bb
     W_{-1,-2}(z,{\bar z}) O_k (w,{\bar w})
       = \frac{1}{z-w} \frac{k^2(q+k)}{4q^3}\sqrt{\frac{q+1}{q}}
            O_{q+k}(w,{\bar w}) ~,
\ee
where we use the notation
\bb
     O_k (z,{\bar z}) = {\bar c}({\bar z})c(z)V_k (z,{\bar z}) ~.
\ee
such that $O_k = \int d^2 z b_{-1}{\bar b}_{-1} \cdot O_k (z,{\bar z})$.
The derivative $\pd_{{\bar z}}$ picks up the OPE singularity and so we
get
\begin{eqnarray}
  &&  0 = \int d^2 z \pd_{{\bar z}} \ll W_{-1,-2}(z,{\bar z})
                \prod_{j\in S}O_j \gg_g
                     \nonumber  \\
  && \quad  = \pi \frac{(2q+1)^2 (3q+1)}{4q^3} \sqrt{\frac{q+1}{q}}
             t \ll O_{3q+1} \prod_{j\in S} O_j \gg_g
                     \nonumber  \\
  && \qquad    - \pi \frac{(q+1)}{4q^3} \sqrt{\frac{q+1}{q}}
             \mu \ll O_{q+1} \prod_{j\in S} O_j \gg_g
                     \nonumber \\
  && \qquad  + \pi \frac{1}{4q^3} \sqrt{\frac{q+1}{q}}
             \sum_{k\in S}k^2(q+k)
                \ll O_{q+k} \prod_{j\in S \atop (j\neq k)} O_j \gg_g
                     \nonumber  \\
  && \qquad + \int d^2 z \ll \pd_{{\bar z}} W_{-1,-2}(z,{\bar z})
                \prod_{j\in S} O_j \gg_g  ~.
\end{eqnarray}
The first and the second correlators of r.h.s. come from the OPE with
the screening charge $O_{2q+1}$ and the cosmological constant $O_1 $
respectively. Usually the last correlator would vanish because the
divergence of the current is the BRST trivial (2.22). However, as discussed
in Sect.3, the boundary of moduli space pinching 2D surface are
dangerous and we have to evaluate it carefully.

  Using the expression of factorization we calculate the following
quantity
\begin{eqnarray}
  && \kappa^2 \frac{Q_M}{2\pi} \sum^{\infty}_{N=0} \sum^{\infty}_{k=1}
      \int^{\infty}_{-\infty} \frac{dp}{2\pi}
         < F_1 ~ \int_{|z| \leq 1} d^2 z
            \pd_{{\bar z}} W_{-1,-2}(z,{\bar z})
           \nonumber  \\
  && \qquad\qquad
         \times D |-p, \beta_{-k}, N >< p, \beta_k, N | F_2 > ~,
\end{eqnarray}
where $D $ is the propagator. $F_1 $ and $F_2 $ are  sets composed
of the operators in $S$ and also the cosmological constants and
the screening charges now. Since the BRST charge commutes with the
Hamiltonian, there is no contribution from $1/H $ term in the
propagator. While the boundary term is singular and there is a possibility
that nonvanishing quantities remain in the limit $\tau \rightarrow
\infty$. So we evaluate
\begin{eqnarray}
  &&   \lim_{\tau \rightarrow \infty} \kappa^2 \frac{Q_M}{2\pi}
         \sum^{\infty}_{k=1} \int^{\infty}_{-\infty} \frac{dp}{2\pi}
           \int_{{\rm e}^{-\tau}\leq |z| \leq 1} d^2 z
          < F_1 ~ [{\bar b}_{-1}, W_{-1,-2}(z,{\bar z})]
              \nonumber  \\
  && \qquad \times  {\bar Q}_{BRST}
         \Bigl( -\frac{2\pi}{H}{\rm e}^{-\tau H} \Bigr)
             |-p, \beta_{-k} > < p, \beta_k | F_2 > ~,
\end{eqnarray}
where we omit $N \neq 0 $ modes because, as discussed below, these
modes vanish at $\tau \rightarrow \infty$. Noting ${\bar {\hat d}}
|-p, \beta_{-k} > ={\bar b}_0 |-p,\beta_{-k}>=0 $ we get
\bb
      {\bar Q}_{BRST} \Bigl( -\frac{2\pi}{H} {\rm e}^{-\tau H}
               \Bigr) |-p,\beta_{-k} >
         = -\pi {\bar c}_0 {\rm e}^{-\tau H} |-p,\beta_{-k}> ~.
\ee
{}From this and the relation $[ {\bar b}_{-1},
{\bar B}_{-1,-2}({\bar z}) ]=-{\bar b}({\bar z})
{\rm e}^{\a_{3q+1}\phi({\bar z})}
{\rm e}^{i\beta_{q-1}\varphi({\bar z})}$, we obtain the expression
\begin{eqnarray}
    \lefteqn{
              \lim_{\tau \rightarrow \infty} \kappa^2 \frac{Q_M}{4\pi}
              \sum^{\infty}_{k=1} \int^{\infty}_{-\infty} dp
              A_{-1,-2} \bigl( i(-p+Q_L), \beta_{-k} \bigr)
              {\rm e}^{-\tau(p^2 -Q^2_L +2\Delta_k -2)}
             }   \nonumber  \\
   & & \times \int_{{\rm e}^{-\tau} \leq |z| \leq 1} d^2 z
           |z|^{ 2\{-2-i\a_{3q+1}(-p+Q_L)
                   -\Delta_{q-1}-\Delta_k +\Delta_{q-k} \} }  \\
   & & \times  < F_1 |-p-i\a_{3q+1}, \beta_{q-k} >
            < p, \beta_k | F_2 > ~,       \nonumber
\end{eqnarray}
where $\Delta_k $ is defined by (3.3) and
\bb
    A_{-1,-2}(\a,\beta)=\frac{1}{q} \biggl[
        -\frac{q-3}{4\sqrt{2}}\a -\frac{7q+3}{4\sqrt{2}}\beta
        +\frac{q^2-2q-1}{8\sqrt{q(q+1)}} (\a^2 +\beta^2)
        +\frac{3q^2+2q+1}{4\sqrt{q(q+1)}} \a\beta \biggr] ~.
\ee
Changing the variable to $z={\rm e}^{-\tau x +i\theta} $, where
$0 \leq x \leq 1 $ and $0 \leq \theta \leq 2\pi $, the above
expression is rewritten as
\begin{eqnarray}
  &&        \lim_{\tau \rightarrow \infty} \kappa^2 \frac{Q_M \tau}{2}
            \sum^{\infty}_{k=1} \int^{\infty}_{-\infty} dp
            \int^1_0 dx A_{-1,-2} \bigl( i(-p+Q_L), \beta_{-k} \bigr)
               \nonumber    \\
  && \quad \times < F_1 |-p-i\a_{3q+1}, \beta_{q-k} >< p, \beta_k | F_2 >
              \exp \Bigl[ -\tau \bigl\{ p^2 -Q^2_L +2\Delta_k -2
                 \bigr. \Bigr.  \nonumber \\
  && \qquad\qquad\qquad  \Bigl. \bigr.  +2x(-i\a_{3q+1}(-p+Q_L)-1
               -\Delta_{q-1}-\Delta_k +\Delta_{q-k} )
               \bigr\} \Bigr] ~.
\end{eqnarray}
Since the exponential term is highly peaked in the limit
$\tau \rightarrow \infty$, the saddle point estimation becomes exact.
The saddle point of the $p$ integral is $p=-i\a_{3q+1} x$, so that
(4.14) becomes
\begin{eqnarray}
    &&       \lim_{\tau \rightarrow \infty}\kappa^2 \frac{Q_M\tau}{2}
             \sqrt{\frac{2\pi}{2\tau}} \sum^{\infty}_{k=0}
             \int^1_0 dx A_{-1,-2}(-\a_{3q+1} x+iQ_L ,\beta_{-k})
             \exp \Bigl( -\tau Q^2_M (qx-k)^2 \Bigr)
                \nonumber  \\
    & & \qquad\qquad \qquad \times < F_1 | i(x-1)\a_{3q+1},\beta_{q-k} >
                <-ix\a_{3q+1},\beta_k |F_2 >  ~.
\end{eqnarray}
The $x$ integral is also evaluated at the saddle point
\bb
      x= \frac{k}{q} ~.
\ee
To give nonvanishing contributions it is necessary that the saddle points
are located within the interval $0 \leq k/q \leq 1$. Thus the sum of
the integer $k$ is restricted within $0 < k \leq q$ and we get
\bb
   \kappa^2 \frac{\pi}{8q^3} \sqrt{\frac{q+1}{q}} \sum^{q}_{k=1}
         k(q-k) < F_1 ~ O_{q-k} >< O_k ~ F_2 > ~.
\ee
Note that the edge state of $k=q $ vanishes because of the factor
$k(q-k)$.

   Replacing $\Delta_k $ and $\Delta_{q-k}$ with $\Delta_k +N$ and
$\Delta_{q-k}+N$ in the expression (4.14), we can see that the
oscillation modes vanish exponentially as ${\rm e}^{-2N\tau}$. Therefore
we obtain
\begin{eqnarray}
   &&        \int d^2 z \ll \pd_{{\bar z}} W_{-1,-2}(z,{\bar z})
             \prod_{j\in S} O_j \gg_g
                \nonumber  \\
   &&  = \kappa^2 \frac{\pi}{8q^3} \sqrt{\frac{q+1}{q}}
             \sum^{q-1}_{k=1} k(q-k) \biggl[
                \ll O_{q-k} O_k \prod_{j\in S} O_j \gg_{g-1} \biggr. \\
   && \qquad \biggl. +\frac{1}{2} \sum_{S=X\cup Y \atop g=g_1 +g_2}
                  \ll O_{q-k} \prod_{j\in X} O_j \gg_{g_1}
                  \ll O_k \prod_{j\in Y} O_j \gg_{g_2} \biggr]
                \nonumber
\end{eqnarray}
The first term of r.h.s. is a variant of the boundary (4.17), where
a handle is pinched.

  Rescaling the normalization of the scaling operator as
\bb
    O_l \rightarrow \frac{q}{l}
            \frac{\Gamma(1+l-l\rho)}{\Gamma(-l+l\rho)} O_l
\ee
and
\begin{eqnarray}
  & & t \rightarrow \frac{2q+1}{q}
       \frac{\Gamma(-2q-1+(2q+1)\rho)}{\Gamma(2q+2-(2q+1)\rho)} t ~,\\
  & & \mu \rightarrow \frac{1}{q}
        \frac{\Gamma(-1+\rho)}{\Gamma(2-\rho)} \mu ~,
\end{eqnarray}
where $\rho=(q+1)/q$, we finally get the equation
\begin{eqnarray}
  && 0=(2q+1)t \ll O_{3q+1} \prod_{j\in S} O_j \gg_g
       +x    \ll O_{q+1} \prod_{j\in S} O_j \gg_g
        \nonumber   \\
  && \qquad + \sum_{k\in S} k \ll  O_{q+k}
              \prod_{j\in S \atop (j\neq k)} O_j \gg_g      \\
  && \qquad +\frac{\lambda}{2} \sum^{q-1}_{k=1}
             \biggl[  \ll O_{q-k} O_k \prod_{j\in S} O_j \gg_{g-1}
                 \biggr.   \nonumber  \\
  && \qquad\qquad\qquad \biggr.
                  +\frac{1}{2} \sum_{S=X\cup Y \atop g=g_1 +g_2}
                  \ll O_{q-k} \prod_{j\in X} O_j \gg_{g_1}
                  \ll O_k \prod_{j\in Y} O_j \gg_{g_2} \biggr]
                \nonumber  ~,
\end{eqnarray}
where we set $x=-\mu $ and $\lambda =-\kappa^2$.
Note that the edge states do not
appear in the expression as far as the operators in $S$ do not
include the edge states. This equation is nothing but the Virasoro
constraint
\bb
     L_1 = \sum_k k x_k \pd_{q+k}
            +\frac{\lambda}{2} \sum_{k+l=q}\pd_k \pd_l
\ee
with $x_1 = x $, $x_{2q+1}=t $ and other $x_j ~ '{\rm s}=0 $.

\subsection{$L_n$ equation}
\indent

   In this subsection we discuss the current $W_{-1,-n-1}$,
which has the form
\bb
     W_{-1,-n-1}(z,{\bar z})
          = w(z,{\bar z}){\rm e}^{\a_{(n+2)q+1}\phi(z,{\bar z})}
              {\rm e}^{\beta_{nq-1}\varphi(z,{\bar z})} ~,
\ee
where $w(z,{\bar z}) $ is the non-exponential part with conformal
dimensions $(n+1,n)$. The OPE of  the current with the screening charge
gives the scaling operator $O_{(n+2)q+1}=\sigma_{n+2}(O_1)$. The
factorization form can be evaluated as done in Sect 4.2 and we get
\bb
     \sum^{nq}_{k=1} w(k) < F_1 ~ O_{nq-k} >
                < O_k ~ F_2 >  ~.
\ee
To calculate the factor $w(k) $ we need to know the explicit form of
$w(z,{\bar z})$. For $n=1 $ we get $w(k) \propto k(q-k)$. In general
$R_{-1,-n-1}(z)$ part of the current gives the $(n+1)$-th order
polynomial of $k$ and the factor $w(k)$ will have the form
\bb
      w(k) \propto k(k-q)(k-2q) \cdots (k-nq) ~.
\ee
Then the edge states decouple and the expression can be rewritten as
\bb
    \sum^n_{s=1} \sum^{q-1}_{r=1} w \bigl( (s-1)q+r \bigr)
         < F_1 ~ \sigma_{n-s}(O_{q-r}) >
         < \sigma_{s-1}(O_r) ~ F_2 >  ~.
\ee
Thus we identify the Ward identity for the current $W_{-1,-n-1}$ with
$L_n $ constraint.

\section{$W$ equations}
\indent

   The $W$ structures are more complicated than the Virasoro
ones. Here we argue mainly the Ward identity for the currents
$W_{-2,-1}$, in which the essence of $W$ structures is included.

  Since the OPE of $W_{-2,-1}(z,{\bar z})$ with
$O_k (0,0)$ is regular, we need to evaluate the OPE
\bb
   W_{-2,-1}(z,{\bar z}) O_k (0,0) \int d^2 w V_l (w,{\bar w})
      =\frac{1}{z} C(k,l) O_{k+l-q}(0,0) ~.
\ee
After contracting all operators the coefficient $C(k,l)$ is calculated
as
\bb
   C(k,l)=A_{-2,-1}(\a_k ,\beta_k)I_1
             +A_{-2,-1}(\a_l ,\beta_l)I_{-1}
               +a_{-2,-1}(k,l)I_0 ~,
\ee
where
\bb
      A_{-2,-1}(\a ,\beta)=\frac{1}{q+1} \biggl[
             \frac{q+4}{4\sqrt{2}}\a - \frac{7q+4}{4\sqrt{2}}\beta
             - \frac{q^2+4q+2}{8\sqrt{q(q+1)}}(\a^2 +\beta^2)
             + \frac{3q^2+4q+2}{4\sqrt{q(q+1)}}\a\beta
                \biggr]
\ee
and
\bb
    a_{-2,-1}(k,l)=A_{-2,-1}(\a_k +\a_l ,\beta_k +\beta_l)
               -A_{-2,-1}(\a_k ,\beta_k)-A_{-2,-1}(\a_l ,\beta_l) ~.
\ee
The integrals $I_n ~(n=0,\pm 1)$ are defined by
\bb
     I_n = \int d^2 y |y|^{2(k+l-2q)/q} |1-y|^{-2l/q}
                       (1-y)^n ~,
\ee
where we introduce the variable $y=w/z$. They are calculated as
(see ref.~\cite{klt} )
\bb
     I_0 =\pi D(k,l;k+l-q) ~, \quad
     I_1=\frac{q-l}{k}I_0 ~, \quad I_{-1}=\frac{q-k}{l}I_0 ~,
\ee
where $D$ function is defined by
\bb
      D(a,b;c)=\frac{\Gamma(1+a-a\rho)\Gamma(1+b-b\rho)
                      \Gamma(-c+c\rho)}{\Gamma(-a+a\rho)
                      \Gamma(-b+b\rho)\Gamma(1+c-c\rho)} ~.
\ee
Thus we get
\bb
     C(k,l)=\frac{\pi}{2\sqrt{q(q+1)}}(k+l-q)D(k,l;k+l-q) ~.
\ee
Note that, if $k+l-q =nq,~n\in {\bf Z}_+$, the coefficient
$C(k,l)$ vanishes because $D(a,b;c)$ vanishes at
$c=nq,~n\in {\bf Z}_+$.

   We also need to calculate the following boundary
\begin{eqnarray}
    && \lim_{\tau \rightarrow \infty}
          \kappa^2 \frac{Q_M}{2\pi} \sum^{\infty}_{k=1}
            \int^{\infty}_{-\infty}\frac{dp}{2\pi}
         < F_1 ~ \Bigl\{ \int_{{\rm e}^{-\tau}\leq |z| \leq 1}d^2 z
              {\bar \pd}W_{-2,-1}(z) \int_{|w| \leq |z|}d^2 w V_l (w)
            \nonumber  \\
   &&\qquad\qquad\qquad   +\int_{{\rm e}^{-\tau}\leq |z| \leq 1}d^2 z V_l (z)
                   \int_{|w| \leq |z|}d^2 w {\bar \pd}W_{-2,-1}(w)
                 \Bigr\}      \\
    && \qquad\qquad\qquad \times
            \frac{-2\pi}{H}{\rm e}^{-\tau H}
              |-p, \beta_{-k} >< p, \beta_k | F_2 > ~.
                 \nonumber
\end{eqnarray}
As done in the previous section the integrals of $p$ and $z$ are
evaluated by using the saddle point method,
\begin{eqnarray}
   && \kappa^2  \frac{\pi}{2(l-q)} \sum^{l-q}_{k=1}
        \bigl\{ A_{-2,-1}(\a_{-k},\beta_{-k}){\tilde I}_0
                +A_{-2,-1}(\a_l,\beta_l){\tilde I}_2
                +a_{-2,-1}(-k,l){\tilde I}_1
         \bigr\}   \nonumber   \\
   && \qquad\qquad\qquad \times
          < F_1 ~ O_{l-k-q} >< O_k ~ F_2 > ~.
\end{eqnarray}
Note that the sum of $k$ is restricted within the interval
$0 < k \leq l-q$,where $l-q >0 $. The integrals ${\tilde I_n}~
(n=0,1,2)$, which comes from the $w$ integrals, are defined by
\bb
     {\tilde I}_n = \int d^2 y |y|^{2(l-k-2q)/q}
                      |1-y|^{2(q-l)/q}
                        \Bigl( \frac{1}{1-y} \Bigr)^n ~,
\ee
where the region $|y| \leq 1$, $y=w/z$ is given by the
first integral of $w$ in (5.9) and the region $|y| \geq 1$, $y=z/w$
is given by the second.  They are calculated
as
\bb
    {\tilde I}_0 =-\pi \frac{(l-q)^2}{q^2}
                     \frac{1}{D(k,l-k-q;l)} ~,
       \quad {\tilde I}_1 =\frac{k}{l-q} {\tilde I}_0 ~,
       \quad {\tilde I}_2 =\frac{k(k+q)}{l(l-q)} {\tilde I}_0 ~.
\ee
Thus we get the expression
\bb
    \frac{-\kappa^2 \pi^2}{4q^2 \sqrt{q(q+1)}} \sum^{l-q-1}_{k=1}
           k(l-k-q) D^{-1}(k,l-k-q;l)
            < F_1 ~ O_{l-k-q} >< O_k ~ F_2 > ~.
\ee
Here also, due to the existence of the inverse of $D$ function and the
factor $k(l-k-q)$, the edge states indeed decouple.

   Combining the boundary contributions (5.1) and (5.13) and rescaling the
fields by using relations (4.19-21)  we get the following Ward identity
\begin{eqnarray}
   && 2(2q+1)^2 t^2 \ll O_{3q+2} \prod_{j\in S} O_j \gg_g
        +2(2q+1)xt \ll O_{q+2} \prod_{j\in S} O_j \gg_g
             \nonumber  \\
   && +2(2q+1)t \sum_{k\in S}k \ll O_{q+k+1}
                  \prod_{j\in S\atop (j\neq k)} O_j \gg_g
            +2x \sum_{k\in S}k \ll O_{k+1-q}
                  \prod_{j\in S\atop (j\neq k)} O_j \gg_g
               \nonumber  \\
   && +2 \sum_{k,l\in S}kl \ll O_{k+l-q}
                  \prod_{j\in S\atop (j\neq k,l)} O_j \gg_g
                     \nonumber   \\
   && +(2q+1)\lambda t \sum^{q-1}_{k=2} \Bigl[
                 \ll O_{q+1-k} O_k \prod_{j\in S} O_j \gg_{g-1}
                  \Bigr.     \\
   &&\qquad\qquad \Bigr.  +\frac{1}{2}
                \sum_{S=X\cup Y \atop g=g_1 +g_2}
                \ll O_{q+1-k} \prod_{j\in X} O_j \gg_{g_1}
                \ll O_k \prod_{j\in Y} O_j \gg_{g_2}
                  \Bigr]  \nonumber \\
   && +\lambda \sum_{l\in S} \sum^{l-q-1}_{k=1} l \Bigl[
                 \ll O_{l-k-q} O_k \prod_{j\in S\atop (j\neq l)}
                     O_j \gg_{g-1}
                  \Bigr. \nonumber  \\
   &&\qquad\qquad \Bigr.  +\frac{1}{2}
                \sum_{S=X\cup Y \atop g=g_1 +g_2}
                \ll O_{l-k-q} \prod_{j\in X\atop (j\neq l)} O_j \gg_{g_1}
                \ll O_k \prod_{j\in Y\atop (j\neq l)} O_j \gg_{g_2}
                  \Bigr]=0 ~, \nonumber \\
\end{eqnarray}
where $\lambda=-\kappa^2$ and $x=-\mu$. The edge states are removed.
The first term  is given by choosing the two screening charges as
$O_k$ and $O_l$ in eq.(5.1). The second corresponds to choosing the
cosmological constant and the screening charge, and so on.
This is nothing but the
$W^{(3)}_{-1}$ constraint described as
\bb
      W^{(3)}_{-1}=\sum_{-l-k+m=-q} l k x_l x_k \pd_m
                   +\lambda \sum_{-l+k+m=-q} l x_l \pd_k \pd_m
\ee
with $x_1 =x$, $x_{2q+1}=t $ and other $x_j ~'{\rm s}=0$.

   In $W^{(3)}_n$ algebra there exists the three derivative term
\bb
         \frac{\lambda^2}{q}\sum_{l+k+m=nq}\pd_l \pd_k \pd_m ~.
\ee
This term can be calculated as a variant of the boundary (5.9) by
replacing $V_l$ with $\kappa^2 (Q_M/2\pi)(1/h_l)V_{-l}<O_l
{}~F_3 >$, where $1/h_l =\int dp ~(p^2 +E_{l,0})^{-1} =\pi/lQ_M$.
For $n=-1$ this term vanishes.

  In general cases it is necessary to calculate the following operator
product
\bb
     W_{-k,-n-k}(z,{\bar z})O_{l_1}(0,0) \int V_{l_2} \cdots \int V_{l_k}
       ~ \propto ~ \frac{1}{z} O_{nq+l_1 +\cdots +l_k}(0,0) ~.
\ee
This OPE corresponds to the single derivative term of $W^{(k+1)}_n$
constraint
\bb
      W^{(k+1)}_n =\sum_{-l_1 -\cdots -l_k +m=nq}
         l_1 \cdots l_k x_{l_1} \cdots x_{l_k} \pd_m ~+~\cdots ~.
\ee
If we take $l_1 = \cdots =l_k =2q+1 $, this produces the operator
$O_{(n+2k)q+k}=\sigma_{n+2k}(O_k)$. We also need to calculate the
the boundary pinching 2D surface for the two derivative term. The
terms with more derivatives are calculated as variants of
the two derivative term.

   Finally we comment on the OPE algebra of the currents
$W_{-k,-n-k}$. We identify the Ward identities of the currents
$W_{-k,-n-k}$ with the $W^{(k+1)}_n$ constraints which form the
$W_q$ algebra. Contrary to this the conservation of the momentums
indicates that the operator algebra of the
currents forms rather the $W_{\infty}$ algebra than the $W_q$ algebra.
Here we  conjecture that in the correlation functions the
$W_{\infty}$ algebra reduces to the $W_q$ algebra and the currents of
$k \geq q$ becomes redundant. The similar argument appears in the
matrix model approach~\cite{fknb}.

\section{Discussion}
\indent

   Until now we considered the interaction theory perturbed by the
cosmological constant $O_1 $ and the screening charge $O_{2q+1}$. The
formalism is easily generalized to the arbitrary potential model,
\bb
    S=S_0 -\sum_j x_j O_j ~.
\ee
If we take $x_1 =-\mu $, $x_{p+q}=t$ and other $x_j ~'{\rm s}=0$, we will
get the $(p,q)$ conformal theory coupled to gravity. Replacing the
screening charge $O_{2q+1}$ with the operator $O_{p+q}$, the
definition of the correlation function changes into
\bb
   \ll \prod_{j \in S}O_j \gg_g
      =\kappa^{-\chi} \mu^s \frac{\Gamma(-s)}{\a_1}
                \frac{2\pi}{Q_M} \frac{t^n}{n!}
          < \prod_{j \in S}O_j ~ (O_1)^s (O_{p+q})^n >_g ~,
\ee
where
\begin{eqnarray}
    & & s = \frac{1}{p+q-1}[(p+q)\chi -\sum_{j \in S}(p+q-j) ] ~, \\
    & & n = \frac{1}{p+q-1}[-\chi +\sum_{j \in S}(1-j) ] ~.
\end{eqnarray}
The Ward identities of this model are easily derived by using the
results (4.6), (4.17), (5.1) and (5.13), where we have only to lay
the roles that the screening charge did on the operator $O_{p+q}$.

   It may be straightforward to generalize the formalism to the
supersymmetric model. The physical states of the superconformal model
coupled to gravity are studied in ref.~\cite{io,bmpb}

   The realization of the nonlinear structures is due to the pole
structure of the propergator. As a result only the states
satisfying the $H=0$ condition survive in the intermediate line. The
$H=0$ condition is the defining-equation of the quantum gravity, so
it seems to be natural that the $H=0$ condition is preserved in the
intermediate line of the quantum gravity.\footnote{Note that the $H=0$
condition is realized in the correlators, but not for the states in
general. For instance $HV_l (1) |{\rm phys}>\neq 0$.}
%%%%%%%%%%%%%
This aspect breaks down for the $c_M > 1$ model couped to 2D
gravity.  Whether it is always realized for the
well-defined quantum gravity in general
is an important question in the future.

   The unitarity is also an important issue in the quantum gravity. In
the Liouville theory approach, from the hermiticy of the Virasoro
algebra $L^{\dagger}_n =L_{-n}$, we can see that the Liouville and the
matter fields have the positive and the negative metric respectively.
Together with the $b$ and $c$ ghosts they form the BRST quartet, so
the no-ghost theorem goes well as in the string theory~\cite{ko}.

   As a nontrivial model there is the 2D quantum dilaton
gravity~\cite{hb,ht}. This model has
the feature that, if the matter's degree of freedom is greater than
the critical value determined by the theory, there appears the region
where two negative metric
fields exist. Naively this region is dangerous in the unitarity and so,
in order that the theory is well-defined, it seems that the matter's degree of
freedom should be restricted. Furthermore  it is shown that, in this case, the
curvature singularity disappears by the quantum effects. It may give
the another interpretation of the Hawking radiation.

\vspace{0.5cm}
\noindent{\bf Acknowlegement:}
  I am very grateful to M. Kato for many valuable discussions.
I also wishes to thank K. Itoh and S. Mizoguchi for helpful
comments and Y. Kazama and M. Ninomiya for encouragement.

\end{document}